\documentclass{nature}

\usepackage{graphicx}
\usepackage{xcolor}
\usepackage{changes}
\usepackage[left=2.0cm, right=2.0cm, top=2.00cm, bottom=2.00cm]{geometry}
\makeatletter
\let\saved@includegraphics\includegraphics
\AtBeginDocument{\let\includegraphics\saved@includegraphics}
\renewenvironment*{figure}{\@float{figure}}{\end@float}
\makeatother

\title{Giant localised spin-Peltier effect due to ultrafast  domain walls motion in antiferromagnetic metals}

\author{R. M. Otxoa$^{1,2,*}$, U. Atxitia$^{3}$, P. E. Roy$^{1}$ $\&$ O. Chubykalo-Fesenko$^{4}$}

\begin{document}

\maketitle

\begin{affiliations}
 \item Hitachi Cambridge Laboratory, J. J. Thomson Avenue, Cambridge CB3 0HE, United Kingdom
  \item Donostia International Physics Center, Donostia San  Sebastian 20018, Spain
 \item Dahlem Center for Complex Quantum Systems and Fachbereich Physik,  Freie Universit\"{a}t Berlin,  14195 Berlin, Germany
 \item Instituto de Ciencia de Materiales de Madrid, CSIC, Cantoblanco, 28049 Madrid, Spain
\end{affiliations}

\begin{abstract}
        Spin thermo-electric phenomena have attracted wide attention  recently, e.g. the Spin Peltier effect (SPE) -- heat generation by magnonic spin currents. Here we find that the Spin Peltier effect also manifests as a heat wave accompanying fast moving magnetic textures.   
High speed and extreme magnetic excitation localisation are paramount for efficient transfer of energy from the spin-degrees of freedom to electrons and lattice. While satisfying both conditions is subject to 	severe restrictions in ferromagnets, we find that domain walls in antiferomagnets can overcome these limitations due to their potential ultrahigh mobility and ultra-small widths originating from the relativistic contraction. To illustrate our findings, we show that electric current driven domain wall motion in the 	antiferromagnetic metal Mn$_2$Au can carry a localised heat wave with the maximum amplitude up to 1 K . Since domain walls are well localised nanoscale magnetic objects, this effect has the potential for nanoscale heating sensing and functionalities. 
\end{abstract}

             Heat production in electronics constitutes one of the major problems of the ever-growing information technology society and is a major source of energy waste. 
             To convert this waste to useful green house technologies is a major societal challenge and a research there is doing its first steps. 
             For instance, nanoscale heating has become crucial for the existence for several emerging disruptive technologies, from cancer treatments, to information technologies such as heat assisted recording\cite{Challener}.  
             Still, it remains a challenge to find ways not only to control the level of heating  but also to do it in a fast, energy-efficient manner at ever decreasing length scales. 
              Spin thermophysics offer a wide plethora of phenomena to be exploited at the nanoscale, such as the Spin Seebeck \cite{UchidaNature2008} and its inverse the Spin Peltier effect \cite{DaimonNatComm2016,UchidaNature2018}, i.e., the generation of thermal gradients with spin currents. The experimental demonstration of the SPE has become possible only recently
              based on the idea that spin currents passing through a magnetic texture would heat one end and cool down the other, 
              thereby creating a thermal gradient. 
              At the same time, ultrafast manipulation and control of spin degrees of freedom is possible by the use of femtosecond laser pulses\cite{KirilyukRMP2010}, and picosecond electric currents\cite{YangScienceAdv2018}, and hold the promise to transform information technology. Here, we theoretically propose that DWs moving at high speed are accompanied by heat-waves localised at the DW.

              In terms of utilising high speed and localised magnetic textures, the idea of racetrack memory architecture, as proposed for fast, high density 3D magnetic memories is very useful.              
               Race track memory concepts are based on moving textures, such as domain walls (DWs)\cite{ParkinScience2008} or skyrmions\cite{FertNatNanotech2013}, localised to move within the confines of a magnetic track. 
              The feasibility of ultrafast race-track memories calls for the use of antiferro or ferri-magnetic systems  \cite{WadleyScience16,YangNatNano2016,BaltzRMP2018,NatPhysReview2018Jungwirth}.  
              Antiferromagnets (AFMs) represent one of the most promising candidates in the quest for faster, energy efficient technologies \cite{BarkerPRL2017,NelePRL2017,OlejnikScienceAdvance18}. 
              Theoretically, the velocity in antiferromagnetic textures is not limited by the so called Walker breakdown\cite{SchryerJAP1974,MouginEPL2007}, and can potentially achieve tens of km/s\cite{YangNatNano2016,GomonayPRL2016}, in direct comparison to hundreds of m/s in their ferromagnetic counterpart.      
              Since antiferromagnets are robust against magnetic fields,  
              electric currents have been proposed as an alternative way to control and manipulate AFM spin dynamics, from domain wall motion\cite{ShinnoPRL2016,GomonayPRL2016} to magnetic switching\cite{WadleyScience16}. 
              However, their absence of net magnetic moment make them challenging to probe reliably.

     Here, we demonstrate the possibility to use electric driven AFM DW motion to generate significant localised heating wave, with a transient electronic temperature raise at least three orders of magnitude larger than in the ferromagnetic case.   
      Notably, by developing a kinetic model where both local and non-local electron, phonon and spin relaxation are included, we are able to identify relevant timescales and extend the existing framework for the AFM DW dynamics onto the  case of non-equilibrium ultrafast dynamics.
      
     We show that the magnetic energy conversion takes place in the ultrafast regime where 
     the DW motion at magnonic velocities implies subpicosecond energy transfer to the electronic
     system. Subsequently, those hot electrons release their excess of energy to the lattice via the electron-phonon coupling, a thermalization process that takes several picoseconds. Finally, lateral heat diffusion transports thermal energy away from the DW position in the hundred picosecond  scale. 
   
  Concerning the use of AFM in nanoheating technologies, the standard way of applying cycling magnetic fields would hardly work  due to the small coupling owing to their vanishing net magnetization. However, relatively low electric currents are able to drive the DWs in AFMs up to magnonic velocities which lie in the elusive range of THz. At those ultrafast spin dynamics, two effects emerge which need to be considered; i) Lorentz contraction of the characteristic lengths due to the relativistic nature of AFM spin dynamics, and ii) ultrafast magnetic energy conversion into heat. While the former has been  investigated in literature for some classes of AFMs\cite{GomonayLoktev2008}, the latter effect remains largely unknown.
  In this work, we reveal the fundamental role of the ultra-high speeds and Lorentzian contraction of DW dynamics in the magnetic energy conversion into dissipation of a moving AFM magnetic texture. 
 In some sense the heat production of electrically driven DW motion is analogous to the Joule heating, when passing an electric current through a resistive material. 
     In our proposal, the role of the moving electrons is played by the DW, which implies that the mechanisms of energy conversion into heat occur at the site of the DW.    
     This would enable a highly localized heat wave source, provided that the DW  moves faster than typical diffusion processes, with the added benefit of that the heat source is movable.
     
In terms of viable materials for this approach, antiferromagnetic metals would appear to be favored over counterpart insulators since in the latter the magnetic energy would  be absorbed into the magnon bath rather than transferred to the electronic degrees of freedom \cite{GomonayJPD2018}. Moreover, magnon relaxation in insulators is a slow process, which could be comparable to the thermal diffusion. 
     In metals however, spin degrees of freedom efficiently coupled to the electronic system\cite{Bhattacharjee2018}. 
     This results for example in femtosecond ($10^{-15}$ s) time scale dynamics of spins, electrons and lattice \cite{DornesNature2019}, opening the door to Petahertz ($10^{15}$ s$^{-1}$) spintronics\cite{PetaHertz2019}.   
This strong coupling between the electronic and spin degrees of freedom provides an ideal benchmark for the possibility of ultrafast  energy transfer from spin to the electron system by pure ultrafast  domain wall dynamics in antiferromagnetic metals.

 \section*{Results} 
 \subsection{Heat generation due to a AFM DW moving at ultrafast speeds.--}

             The response of a magnet to an external stimulus strongly depends on  the way dissipation takes place, which in turn controls the magnetic response in processes such as DW motion, switching and spin transport. The dynamics of  the angular momentum dissipation in magnets is well described by the Gilbert relaxation term (see Methods), whose rate, is proportional to the so-called Gilbert damping, $\alpha$. This is connected to the magnetic energy dissipation described by  
             the Rayleigh dissipation functional, $\mathbf{\dot{Q}}_{\rm{DW}}= \int dV \eta \mathbf{\dot{S}}^2(\mathbf{x})$, where $\eta=\mu_{\rm{sl}} \alpha/\gamma$, $\gamma$ is the gyromagnetic ratio and $\mu_{\rm{sl}}$ is the sublattice magnetisation. We use the Rayleigh dissipation functional to            
             estimate the parameter dependence of the initial temperature rise while the DW is moving.  
             Within the Lagrangian formalism for a stationary moving (along $x$-direction) 1D  domain wall in a layered antiferromagnet (or ferromagnet), the Rayleigh dissipation function (per atomic spin) can be derived as (see Supplemental Material) 
           \begin{equation}
                  \dot{Q}_{\rm{DW}}(x,t) =\frac{\mu_{\rm{sl}}}{\gamma} \alpha \left( \frac{v_{\rm{DW}}}{\Delta_{\rm{DW}}}\right)^2\frac{1}{\cosh^{2} (q)},
                  \label{eq:dissipation-analytical}
           \end{equation}
          where $q=(x-v_{\rm{DW}}t)/\Delta_{DW}$, $v_{\rm{DW}}$ is the DW velocity and $\Delta_{\rm{DW}}$ is the DW width. For a stationary moving DW in the absence of thermal diffusion,
          the temperature profile accompanying the DW propagation can be estimated as
            \begin{equation}
                  \Delta T(x,t) =  \frac{2}{C}\frac{\mu_{\rm{sl}}}{\gamma} \alpha \frac{v_{\rm{DW}}}{\Delta_{\rm{DW}}}\tanh(q).
             \label{eq:DT-analytical}
            \end{equation}
            where $C$ is the electron (phonon) bath heat capacity.
           We note that the temperature rise scales with the ratio $v_{\rm{DW}}/\Delta_{\rm{DW}}$, which is advantageous for AFM DW due to the Lorentz contraction of $\Delta_{\rm{DW}}$ down to the atomic limit, and the 
           possibility to achieve magnonic velocities.

 In the following we explore the non-equilibrium (non steady) dynamics of DW in the metallic AFM, Mn$_2$Au, driven by spin-orbit fields in a track and its impact on the track temperature.  
  Efficient DW motion can be achieved in certain AFM metals by injecting electric currents into them. In particular, crystals with
          locally broken inversion symmetry at the magnetic sites A and B form inversion partners, the inverse spin galvanic effect produces a staggered local spin accumulation with opposite polarities. The effect of the current is then to produce local staggered spin-orbit (SO) field which is perpendicular to the spin-polarised current direction and is linearly proportional to its magnitude. The torque generated on each AFM sublattice has, therefore, the same form as in ferromagnets.  Together with the AFM metal CuMnAs, 
           Mn$_2$Au is one of the few AFM materials with high enough critical temperature (1500 K), and  the required crystal symmetry.  
           This makes it suitable for spintronic applications. 
          Moreover, recent estimations of the effective SO fields, $H_{\rm{SO}}$ are  $\sim$ 2 mT per 10$^7$ A/m$^2$. 
          
\begin{figure}
\center
\includegraphics[scale=0.8]{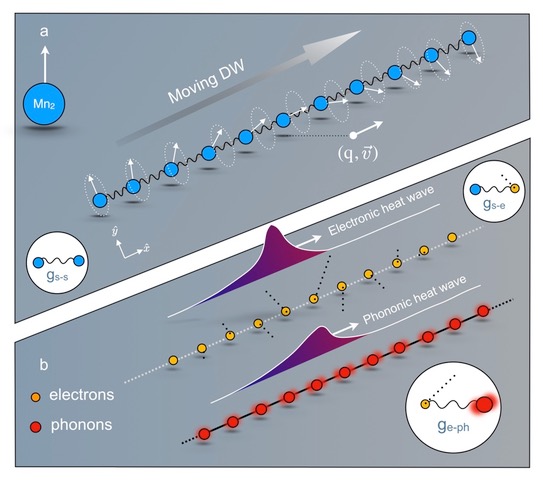}
\caption{\textbf{Spin dynamics, and electron and phonon energy dynamics.} 
(a) In  Mn$_2$Au the ions of Mn provide the spins (white arrows). 
   Two magnetic domains are separated by a moving domain wall (DW), defined by its position and velocity $(q,v)$.  
(b) Subpicosecond heat dissipation is provided via coupling of the spins' degrees of freedom located at the DW to the electrons defined by $g_{\rm{s-e}}$.
This creates a solitonic-like heat wave of hot electrons which accompany the moving DW profile. 
This hot electrons, are now in non thermal equilibrium with the lattice, described by a phonon temperature.  
The electron-phonon coupling, $g_{\rm{e-ph}}$, is responsible for the heat transfer from the hot electrons to the colder phonons. This process thermalize the electron and phonon system to a final common temperature.}
\label{fig:fig1}
\end{figure}

           For the description of the energetics of Mn$_2$Au we consider a classical spin Hamiltonian 
           (See Methods). Within this model,  
           the Mn spins are responsible for the magnetic properties.
           The electrically driven DW dynamics is calculated using   
           computer simulations based on atomistic spin dynamics\cite{Evans} based on the solution of the Landau-Lifshitz-Gilbert (LLG) equation on a discrete lattice. 
                      A moving DW is characterised by its instantaneous position and velocity of its centre, $(q,v_{\rm{DW}})$ (see Fig. \ref{fig:fig1}a). In AFMs, the Lorentz contraction means that the width $\Delta_{\rm{DW}}$ of the DW depends on $v_{\rm{DW}}$ and its limited by the maximum magnon group velocity $v_g$, i.e. $\Delta_{\rm{DW}}=\Delta_0\sqrt{1-(v_{\rm{DW}}/v_g)^2}$  where $\Delta_0$ is the DW width at rest. 
                        An electric current passing through  Mn$_2$Au creates a staggered spin-orbit torque $H_{\rm{SO}}$ which drives the DW at a velocity, 
           $v_{\rm{DW}}= (\gamma/\alpha)H_{\rm{SO}}\Delta_{\rm{DW}}$.  
             Due to energy conservation, a stationary moving DW  dissipates energy into the medium at the same rate as Zeeman energy lowers due to the domain switching. The dynamics of the redistribution of this excess of energy into the different subsystems is the main result of this work.

         Atomic spin dynamics simulations (ASD) permit us to calculate each of the $\dot{\mathbf{S}}_i(t)$, and consequently $q(t)$ and $v_{\rm{DW}}(t)$ and $\Delta_{\rm{DW}}$ are obtained. Those time-dependent quantities are then fed into Eq. 1  to calculate the instantaneous local  $\dot{Q}_{\rm{DW}}(\mathbf{x}_i,t)$. This quantity enters into the equation of motion for the electron and lattice temperature dynamics, which is described by the two temperature model described in Methods section.
       Figure \ref{fig:fig1}b depicts the transient dynamics of the local electron and phonon temperature profiles due to a fast moving DW. The electron temperature shows a peak temperature lagging slightly behind the DW centre. 
            This is due to the direct coupling between the electron and spins, (g$_{\rm{s-e}}$)\cite{Bhattacharjee2018}. 
            At the same time the phonon temperature shows a much smoother profile, 
            owning to the indirect coupling to the heat source (moving DW) via the electron system, (g$_{\rm{e-ph}}$). 
            The  heat wave is well localised around the centre of the DW.     
           The excess of energy in the electronic system is rapidly transferred to the lattice via the electron-phonon coupling at characteristic timescales of the order $ \sim 1$ ps.
           At the same time, lateral heat transport is also present, flow of energy from hot to cold regions.       
                 We should note that additional channels of energy conversion also exists. For instance, magnon creation, which in turn  can transport energy away from the heat source. In our simulations we do not see significant spin wave creation, probably due to the conditions we are assuming here: low temperature and l15 ps linear ramp time of driving current as opposed to a square pulse.  
          Furthermore, we neglected the Joule heating contribution in our model, although its contribution may be larger than those related to the DW motion. However, the Joule heating only provides a homogeneous background. We hope that these effects would be possible to distinguish by proper calibration in the real experimental set up.

           In our simulations, we start by injecting an electric current with the time profile illustrated in Fig. 2a; a rising time of 15 ps up to a peak value of $H_{\rm{SO}}(t)=60$ mT. The value of this field is kept constant for 5 ps before reducing it to zero, with a falling time of 15 ps and 3.5 ps ($H_{\rm{SO}}=0$), we repeat this pattern but with negative values of $H_{\rm{SO}}$. This cycle is run four times. From our ASD simulations  we gain information about the dynamics of the DW width, $\Delta_{\rm{DW}}$ (Figure 2b) as well as the DW velocity, $v_{\rm{DW}}$ (Figure 2c).
           In Figure 2b, we clearly observe that as a result of the Lorentz contraction, the wall length, $\Delta_{\rm{DW}}$, periodically shrinks and expands comprising values from circa 20 nm (DW at rest) down to almost 4 nm.  
           Blue dots indicate time events when the DW width is maximum, which corresponds to minimum DW velocity (see Figure  2c). On the contrary, yellow dots represent a fully compressed DW propagating at speeds close to $v_g$
           Yellow points indicate time moments where the DW width is minimum, which in turn corresponds to maximum DW velocity.
           During the rise-time of $H_{\rm{SO}}(t)$ (refer to \textbf{a}), the DW speed increases almost up to $v_g$ (maximum magnon group velocity), while during the fall-time, the DW speed decreases down to zero. The same process is repeated with negative values of $H_{\rm{SO}}$, thus, the DW repeats the dynamics described before and goes back to the initial position. This periodic cycling is repeated 4 times in our computer simulations, see Figure 2d. During each half-cycle DW displaces 1500 nm (See Supplementary material).  
                        
\begin{figure}
\center
\includegraphics[scale=0.78]{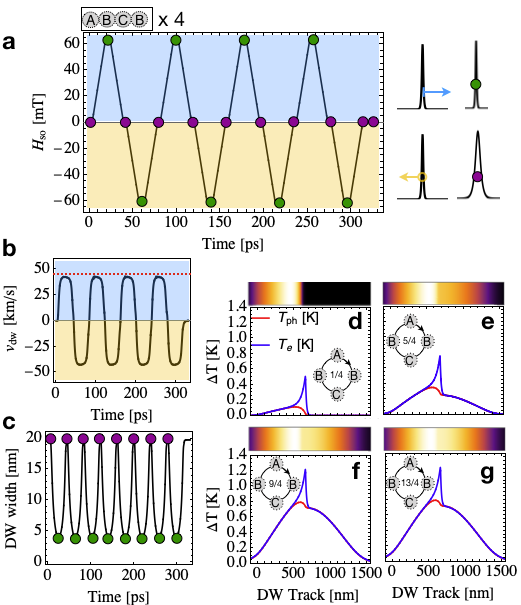}
\caption{ \textbf{Dynamics of the DW driven by spin-orbit fields generated by  electric currents.} \textbf{a} Time dependence of the spin orbit field, $H_{\rm{SO}}(t)$.  \textbf{b} Time resolved $v_{\rm{DW}}$, dashed red lines correspond to $v_g$, the maximum magnon group velocity. \textbf{c} Time resolved DW width, $\Delta_{\rm{DW}}$, where  
purple dots indicate time moments where $\Delta_{\rm{DW}}$ is maximum, 
which corresponds to minimum DW velocity, $v_{\rm{DW}}$  (see (\textbf{b} panel). 
Green dots indicate time moments where $\Delta_{\rm{DW}}$ is minimum ( maximum $v_{\rm{DW}}$).
\textbf{d-g} Spatio-temporal snapshots of the temperature dynamics $\Delta T(t)$ in the sample region where the DW  moves at the time moments corresponding to (d) 1/4 of the period (e) 5/4 of the period (f) 9/4 of the period (g) 13/4 of the period. 
}
\label{fig:fig2}
\end{figure}

           Our main result  is the spatial-temporal dynamics of the temperature increase $\Delta T$ in the sample region where the DW  moves, see Figure \ref{fig:fig2}e-h. The global maximum temperature increase is located in the center of the region of the DW movement (where the velocity is maximum), corresponding to a region of around 500 nm width.
  A snapshot of the spatial profile of the electron and phonon temperature in the track during one DW motion shows clearly the creation of electronic and phonon localised waves with an electronic peak at the DW position, see Figure \ref{fig:fig2}e. The electronic temperature peak at time-scales corresponding to a one driving-cycle and thus accumulates to almost 0.8 K , see Figure \ref{fig:fig2}f. The lateral diffusion is not strong enough to delocalise the heat wave at this timescale. Away from the domain wall position the electron and phonon temperature equilibrate.

 The temperature evolution at the centre of the track (Figure 2 ) shows how the DW motion releases magnetic energy into the electron system in time scales of subpicoseconds, with peak electron temperature appearing only a few hundreds of femtoseconds after the DW passes. This excess of energy is thereafter transferred to the phonon system via electron-phonon coupling in time scales of the order of the picosecond. 
Once the DW has passed lateral thermal diffusion sets in with time scales of the order of hundreds of picoseconds.  These timescales are much shorted than the expected timescale for the energy equilibration with the outside world (the air, for example).

\begin{figure}
\includegraphics[scale=0.8]{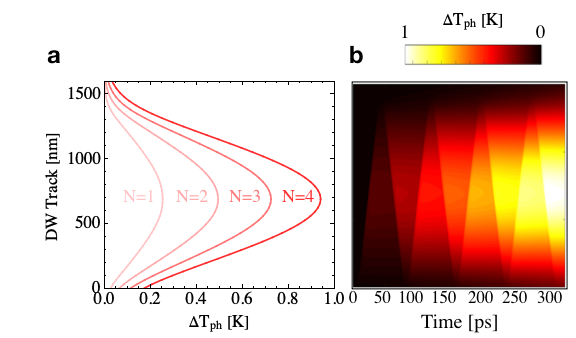}
\caption{\textbf{Heat accumulation in the stripe}. \textbf{a} Spatial phonon temperature $\Delta T_{\rm{ph}}$ distribution after each cycle of the SO field complete cycles, $N=1,2,3,4$, \textbf{b} Spatio-temporal evolution of the accumulated heat in the system.
}
\label{fig:fig3}
\end{figure}

We discuss now the total heat dissipated in our system per cycle (Figure \ref{fig:fig3}a). This excitation protocol produces a phonon heat accumulation at the track centre. In terms of the phonon (electron) temperature the value of 1 K is reached already in 300 ps (Figure \ref{fig:fig3}b) with 4 field cycles with a full width at half maximum of 1 micrometer (Figure \ref{fig:fig3}a). The rapid accumulation is possible  due to the high-speed character of the AFM DWs. At this timescales the heat transfer to the outside media is expected to be small so that a giant magnetocaloric effect is induced.

\section*{Discussion}

The above described mechanism opens the door to  control heating at the nanoscale in a fast manner. 
It is worth discussing the differences and similarities between 
the  heating process by AFM DW and that of  ferromagnetic nanoparticles performing coherent magnetisation rotation by external oscillating magnetic fields. The latter is a standard way to heat, for example, tumor cells in magnetic hyperthermia treatment. 
Energetic considerations show that the magnetic energy density release is different to zero for irreversible processes only, and is equal to the hysteresis loop area.
$\Delta \varepsilon = \int m(H) H dH$, where $m$ is the magnetisation density.
In the absence of other heat losses, one can estimate the maximum temperature rise as $\Delta T \sim \Delta \varepsilon/C$ where $C$ is the specific heat density. 
Although useful, this argument can produce an impression that a relatively large temperature rise can be achieved by only increasing the applied magnetic field. 
The real situation is, however, more complex since dynamical considerations need to be taken into account. Specifically, the rotational speed is in the nanoseconds range, i.e. the maximum possible heat will correspond to fields with the magnitude of the coercive field applied at GHz frequencies. This timescale is much slower than the AFM dynamics considered here. 
Faster field cycling implies minor hysteresis loops leading to a huge decrease of the heating output.
Additionally, at this timescale, the interface heat transfer (to the outside media) is also more efficient than in our case. For comparison, simple estimations show that a small magnetite nanoparticles of 10 nm diameter, under the best conditions of major hysteresis loop would heat up to 10 mK per field cycle.  This estimation is two order of magnitude lower than our calculations for AFM DW motion.

We note that for commonly used ferromagnetic materials, heating by moving DWs would not be so efficient as in AFM. 
For example, for standard parameters of permalloy, we estimate that the DW motion can carry electron temperature pulses of 
a maximum of 1 mK. 

Thermal diffusion also plays an important role in the delocalisation of the temperature rise.  
The heat diffusion rate is defined  by the parameter $\eta=v_{\rm{DW}} \Delta_{\rm{DW}} /D$ where $D=k_e/C_e$ is the electron diffusivity coefficient. 
Considering the typical metal value $D=10^{-4}$ m/s$^{2}$, the above parameter $\eta\ll 1$ for permalloy and $\eta>1$ for Mn$_2$Au. 
Thus the negligible heat wave accompanying the permalloy DW will be completely delocalised, whereas 
for Mn$_2$Au it is localised at the DW.   
 
 Importantly, since eventually heat will spread all over the sample, the particular dynamics of the relaxation of the localised temperature rise is paramount for devising experiments and devices able to exploit this new concept we present here. 
The effect  hinges to the field of ultrafast spin caloritronics  fostered by the recent demonstration of subpicosecond spin Seebeck effect\cite{SeifertNatComm2018}. 
Along this line, it remains a challenge to drive AFM DWs into magnonic velocities, and to develop experimental techniques in order to detect DW dynamics at these ultrafast timescales.
Notably, we believe that nanoscale confined heating at the domain wall position can be used to track the DW position and velocity by measuring the temperature increase with, for example, scanning thermal miscroscopy. Recent  reports indicate that by using the anomalous Nerst effect, thermal nanoscale detection of the DW motion (although in the timescale much smaller than our) is possible \cite{Krzysteczko}.

Interestingly, the effect is not restricted to moving DWs in AFMs but is a universal characteristic of magnetic textures moving at high velocities. Therefore, the concept can be extended to other textures, such as skyrmions or vortices. As example, DWs in cylindrical nanowires and nanotubes lack the Walker breakdown phenomenon  and reach velocities similar to those considered here\cite{YanAPL2011}. Systems with perpendicular anisotropy and Dzyaloshinki-Moriya interactions\cite{Sethi} may also be good candidates to observe the predicted effect. However, the AFM has additional advantage of very small DW width, due to Lorentz contraction at high velocities, which can be reached using SO torques with reasonable current intensities.

\newpage

\begin{methods}

\subsection{Atomic spin structure of Mn$_2$Au.} 
We use a realistic unit cell spin structure for the Mn$_2$Au, see Ref. \cite{RoyPRB2016} for more details. 
The total energy is described with the following atomistic classical spin Hamiltonian, 
     \begin{eqnarray}
      \mathcal{H} &=& - \sum_{i\neq j} J_{ij} \mathbf{S}_{i} \mathbf{S}_{j} 
                    - K_{2\bot} \sum_i \left(\mathbf{S}_{i}\cdot   \mathbf{u}_{3}\right)^2 
                     \nonumber \\ 
                    &  & 
                   - \frac{K_{4\bot}}{2} \sum_i \left(\mathbf{S}_{i}\cdot   \mathbf{u}_{3}\right)^4 
                   - \frac{K_{4\|}}{2} \sum_i \left(\mathbf{S}_{i}\cdot   \mathbf{u}_{2}\right)^4 \nonumber \\ 
                   & & -\mu_{\rm{sl}} \sum_i \mathbf{S}_{i} \cdot \mathbf{H}_i^{\rm{SO}}
                   \label{eq:spin-Hamiltonian}
     \end{eqnarray}
where $\mathbf{S}_i$ is a classical vector,$|\mathbf{S}_i|=1$. The exchange coupling between spins at sites $i$ and $j$, $J_{ij}$ are given by $J_{i1}=-396k_B$, $J_{i2}=- 532 k_B$, and $J_{i3} = 115 k_B$, where $k_B$ is the Boltzmann constant.

Atomic magnetic moment of Mn (magnetically active) atomic spins are $\mu_{\rm{at}}=4\mu_B$, where $\mu_B$ is the Bohr magneton. 
 The biaxial basal-plane anisotropy, $K_{4\|}$ is 1.86$\cdot 10^{-25}$J . The perpendicular anisotropy constants, $K_{2\bot}$= - 1.3$\cdot 10^{-22}$J and  $K_{4\bot}= 2 K_{4\|}$. The in-plane anisotropy to stabilise the domain wall is $K_{2\|}$ = 7 x $K_{4\|}$. Finally, the effect of current due to the spin-orbit interaction of the electron and spins is mapped into an effective staggered magnetic field, $\mathbf{H}_i^{\rm{SO}}$ \cite{RoyPRB2016}. 

\subsection{Atomistic spin dynamics.} 
 The dynamics of the spins in Mn$_2$Au is described by a classical spin model framework based on the phenomenological atomic Landau-Lifshitz-Gilbert equation of motion \cite{Landau1935,Gilbert2004} 
  \begin{eqnarray}  
           (1+\alpha^2){\dot{\mathbf{S}_i}} =
              - \gamma  \mathbf{S}_i  \times  \left( \mathbf{H}_i^{\rm{eff}} +\alpha \mathbf{S}_i  \times \mathbf{H}^{\rm{eff}}_i\right),
             \label{eq:macro-LLG}
\end{eqnarray}
here, $\mathbf{H}^{\rm{eff}}_i=-\frac{1}{\mu_{\rm{sl}}}\partial \mathcal{H}/\partial \mathbf{S}_i$, where  $\mathcal{H}$ is given in Equation \ref{eq:spin-Hamiltonian}. 
Here, $\alpha$ is the magnetic damping parameter, also known as Gilbert damping parameter. The gyromagnetic ratio is $\gamma=1.76\times 10^{11}$ (Ts)$^{-1}$.
Similar to the Hamiltonian, the spin vectors are classical, constant length and normalized, $|\mathbf{S}_i|=1$. 
In the presence of an electric current, the SO coupling creates an effective field, $\mathbf{H}^{\rm{SO}}$, which is added to the effective field, 
$\mathbf{H}^{\rm{eff}}_i$.

\subsection{Two-temperature model} 

            We use the two temperature model to describe the dynamics of the electron and lattice temperature, 
            \begin{eqnarray}
                  C_{\rm{e}} \frac{d T_{\rm{e}}}{d t}&=&g_{\rm{e-ph}} (T_{\rm{e}}-T_{\rm{ph}})+ k_e\frac{\partial^2 T}{\partial x^2} +\dot{Q}_{\rm{DW}-e}(x,t) 
                  \label{eq:electronTmaintext}
                   \\
                  C_{\rm{ph}} \frac{d T_{\rm{ph}}}{d t}&=&-g_{\rm{e-ph}} (T_{\rm{e}}-T_{\rm{ph}}) +\frac{T_{\rm{ph}}-T_0}{\tau_d},    
            \label{eq:phononT}
            \end{eqnarray}
            where $C_{\rm{e}}$, $C_{\rm{ph}}$ are the electron and phonon specific heats, respectively. The coupling between electron and lattice systems is defined by the electron-phonon coupling constant, $g_{\rm{e-ph}}$.
           Lateral heat transport is defined by $k_e\partial^2 T/\partial x^2$, where $k_e$ is the electronic thermal conductivity and $\tau_d$ is the heat difusion time. 
            Exact values used in our model can be found in Table 1.

\end{methods}

\begin{table}
\begin{center}
\begin{tabular}{|c|c|}
\hline
\textbf{Symbol}                & \textbf{Value}                                                                                                                                                                                                                                      \\ \hline
Diffusion coefficient, $k_{\rm{e}}$                       & $200 $                                                                                                                                                                                                                                   \\ \hline
electron heat capacity, $C_{\rm{e}}$                    & $10^{3} \, J/(m^{3} K)$                                                                                                                                                                                                                                       \\ \hline
phonon heat capacity, $C_{\rm{ph}}$                          & $1.5 \cdot 10^{6} \, J/(m^{3} K)$                                                                                                                                                                                                                                   \\ \hline
electron-phonon coupling, $g_{\rm{e-ph}}$                    & $0.25\cdot 10^{18} \,W/ (m^{3} K)$                                                                                                                                                                                                                                             
\\ \hline
Gilbert damping, $\alpha$                               & $0.001$                
\\ \hline                                                                                                                                                                                                                 \end{tabular}
\end{center}

\caption{\textbf{Spin thermophysical parameters of Mn$_2$Au.} Literature values for material parameters relevant for modelling the spin dynamics and heat transfer\cite{StojanovicPRB2010}.}
\label{table:1}
\end{table}



\begin{addendum}
 \item  This work was partially supported by a STSM Grant from the COST Action CA17123.  FU Berlin support by the Deutsche Forschungsgemeinschaft (DFG) through SFB/TRR 227  "Ultrafast Spin Dynamics", Project A08 and the Spanish Ministry of Economy and Competitiveness under grants MAT2016-76824-C3-1-R and FIS2016–78591-C3-3-R  are gratefully acknowledged. 
 \item[Competing Interests] The authors declare that they have no
competing financial interests.
 \item[Correspondence] Correspondence and requests for materials
should be addressed to Rub\'{e}n M. Otxoa .~(email: ro274@cam.ac.uk).
\end{addendum}


\newpage

\newpage

\end{document}